\documentclass[a4paper,12pt]{article}
\usepackage{amsmath, epsfig}

\title{A Common Market Measure for Libor and
Pricing Caps, Floors and Swaps in a Field Theory of Forward Interest Rates}
\author{Belal E. Baaquie
\\Department of Physics, National University of Singapore\\phybeb@nus.edu.sg}
\begin{document}
\maketitle
\begin{abstract}
The main result of this paper that a martingale evolution can be
chosen for Libor such that all the Libor interest rates have a
common market measure; the drift is fixed such that each Libor has
the martingale property. Libor is described using a field theory
model, and a common measure is seen to be emerge naturally for such
models. To elaborate how the martingale for the Libor belongs to the
general class of numeraire for the forward interest rates, two other
numeraire's are considered, namely the money market measure that
makes the evolution of the zero coupon bonds a martingale, and the
forward measure for which the forward bond price is a martingale.
The price of an interest rate cap is computed for all three
numeraires, and is shown to be numeraire invariant. Put-call parity
is discussed in some detail and shown to emerge due to some
non-trivial properties of the numeraires. Some properties of swaps,
and their relation to caps and floors, are briefly discussed.

\end{abstract}

\section{Introduction}
Libor (London Inter Bank Overnight Rates) are the interest rates for
Eurodollar deposits. Libor is one of the main instruments for
interest rates in the debt market, and is widely used for
multifarious purposes in finance. The main focus of this paper is on
the properties of Libor, and in particular finding a common measure
that yields a martingale evolution \cite{MR} for all Libor. Two
other numeraires for the forward interest rates are also considered,
namely the money market numeraire and the forward measure for bonds.

All calculations are performed using the field theory for the
forward interest rates that has been introduced in
\cite{Baaquiecup,Baaquie,Baaquiesv}. The main advantage of modelling
the forward interest rates using field theory is that there are
infinitely many random variables at each instant driving the forward
rates. In particular, for the case of Libor rates, it will be shown,
unlike the usual models in finance, a numeraire can be chosen so
that \textbf{all} the Libor instruments simultaneously have a
martingale evolution \cite{BGM}.

The price of any financial instrument in the future has to be
discounted by a numeraire to obtain its current price. The freedom
of choosing a numeraire results from the fact that for every
numeraire there is a compensating drift such that the price of any
traded instrument is independent of the numeraire.  'Numeraire
invariance' is an important tool in creating models for the pricing
of financial instruments \cite{MR}, and is verified by using three
numeraires for pricing an interest caplet. As expected, the price of
the caplet is numeraire invariant.

In Section 2 the field theory of forward rates is briefly reviewed.
In Section 3 the three numeraires are discussed, and the
corresponding drift velocities are evaluated. In Section 4 the price
of a mid-curve interest caplet is priced for the three numeraires,
in Section 5 put-call parity is derived for the three cases, in
Section 6 interest swaps are discussed, and with some conclusion
drawn in Section 7.

\section{Field Theory Model of Forward Interest Rates} \label{QF}
The field theory of forward rates is a general framework for
modelling the interest rates that allows for a wide choice of
evolution equation for the interest rates.

The Libor forward interest rates $f(t,x)$ are the interest rates,
fixed at time $t$, for an instantaneous loan at future times
$x>t$.\footnote{Libor forward interest rates carry a small element
of risk that is not present in the forward rates that are derived
from the price of zero risk US Treasury Bonds. All calculations in
this paper are based on Libor rates.} Let $A(t,x)$ be a two
dimensional field driving the evolution of forward rates $f(t,x)$
through time, defined by
\begin{equation}
\label{fft}
\frac{\partial f(t,x)}{\partial
t}=\alpha(t,x)+\sigma(t,x)A(t,x)
\end{equation}
where $\alpha(t,x)$ is the drift of the forward interest rates that
will be fixed by a choice of numeraire, and $\sigma(t,x)$ is the
volatility that is fixed from the market \cite{Baaquiecup}. One is
free to choose the  dynamics of how the field $A(t,x)$ evolves.

Integrating eq. \ref{fft} yields
\begin{eqnarray}
\label{frcfrwd}
 f(t,x)=f(t_0,x)+\int_{t_0}^t dt' \alpha(t',x)+\int_{t_0}^t
dt'\sigma(t',x)A(t',x)
\end{eqnarray}
where $f(t_0,x)$   is the initial forward interest rates term
structure that is specified by the market.

The price of a Libor Bond, at present time $t$, that matures at some
future time $T>t$ is denoted by $B(t,T)$, and is defined in terms of
the forward interest rates as follows.
\begin{eqnarray}
B(t,T)=e^{-\int_t^T dx f(t,x)}
\end{eqnarray}

Following Baaquie and Bouchaud \cite{baaqbou}, the Lagrangian that
describes the evolution of instantaneous Libor forward rates is
defined by three parameters $\mu, \lambda, \eta$, and is given
by\footnote{More complicated nonlinear Lagrangians have been
discussed in \cite{Baaquiecup,Baaquiesv}}
\begin{equation}
\label{LA}
{\cal L}[A] = -\frac{1}{2}
\left\{A^2(t,z)+\frac{1}{\mu^2}  \left( \frac{\partial
A(t,z)}{\partial z} \right)^2 +\frac{1}{\lambda^4}
  \left( \frac{\partial^{2}A(t,z)}{\partial^{2}z} \right)^2\right\}
\end{equation}
where market (psychological) future time is defined by
$z=(x-t)^\eta$.

The Lagrangian in  eq. \ref{LA} contains a squared Laplacian term
that describes the stiffness of the forward rate curve. Baaquie and
Bouchaud \cite{baaqbou} have determined the empirical values of the
three constants $\mu, \lambda, \eta$, and have demonstrated that
this formulation is able to accurately account for the phenomenology
of Libor interest rate dynamics. Ultimately, all the pricing
formulae for caps and floors depend on 1) the volatility function
$\sigma(t,x)$, 2) parameters $\mu, \lambda, \eta$ contained in the
Lagrangian, and lastly 3) on the initial term structure.

The action $S[A]$ and the partition function $Z$ of the Lagrangian
is defined as
\begin{eqnarray}
S[A]&=&\int_{t_0}^{\infty} dt \int_0^{\infty} dz {\cal L}[A]\\
Z&=&\int DA e^{S[A]}
\end{eqnarray}
where the symbol $\int DA$ stands for a path integral over all
possible values of the quantum field $A(t,x)$.

All expectation values, denoted by $E[..]$, are evaluated by
integrating over all possible values of the quantum field $A(t,z)$.
The quantum theory of the forward interest rates is defined by the
generating (partition) function \cite{Baaquiecup} given by
\begin{eqnarray}
Z[J]&=& E\big[e^{\int_{t_0}^{\infty} dt \int_0^\infty dz
J(t,z)A(t,z)}\big]\nonumber\\
 &\equiv& \frac{1}{Z}\int DA \, \, e^{S[A]+\int_{t_0}^{\infty} dt \int_0^\infty
dz
J(t,z)A(t,z)}\nonumber\\
\label{zj}
 &=& \exp\Big(\frac{1}{2}\int_{t_0}^{\infty} dt \int_0^{\infty} dz dz'
 J(t,z)D(z,z';t)J(t,z')\Big)
\end{eqnarray}

All financial instruments of the interest rates are obtained by
performing a path integral over the (fluctuating) two dimensional
quantum field $A(t,z)$. The expectation value for an instrument, say
$L[A]$, is defined by the functional average over all values of
$A(t,z)$, weighted by the probability measure $e^S/Z$; the following
notation will be used for denoting the expectation value
\begin{eqnarray}
E\big[L[A]\big]\equiv \frac{1}{Z}\int DA ~L[A]~e^{S[A]}
\end{eqnarray}
This a key equation that relates the formulation of finance based on
stochastic calculus \cite{sc} to the one based on path integrals
\cite{Baaquiecup}; both formulations evaluate the same expectation
values using different formalisms -- in the path integral approach
the averaging is carried out by performing an infinite dimensional
functional integration.

For simplicity of notation, we only consider the case of $\eta=1$
and replace all integrations over $z$ with those over future time
$x$.

\section{Numeraire and Drift} \label{numeraire} The drift velocity
$\alpha(t,x)$ is fixed by the choice of numeraire. The Libor market
measure is first discussed, and then the forward measure and money
market measure are discussed to elaborate different choices for the
numeraire of forward rates, and the drift velocity for each is then
evaluated.

\subsection{Libor Market Measure}
For the purpose of modeling Libor term structure, it is convenient
to choose an evolution such that \textbf{all} the Libor rates have a
martingale evolution. The deposit and payment dates are pre-fixed at
90-day intervals, denoted by $T_n$. The Libor forward interest
rates, denoted by $L(t,T_n)$, are simple interest rates, agreed upon
at time $t<T_n$, for the payment that one would receive for a future
time deposit from $T_n$ to $T_n+\ell$, with  payments made in arrear
at (future) time $T_n+\ell$.

\begin{figure}[h]
  \centering
  \epsfig{file=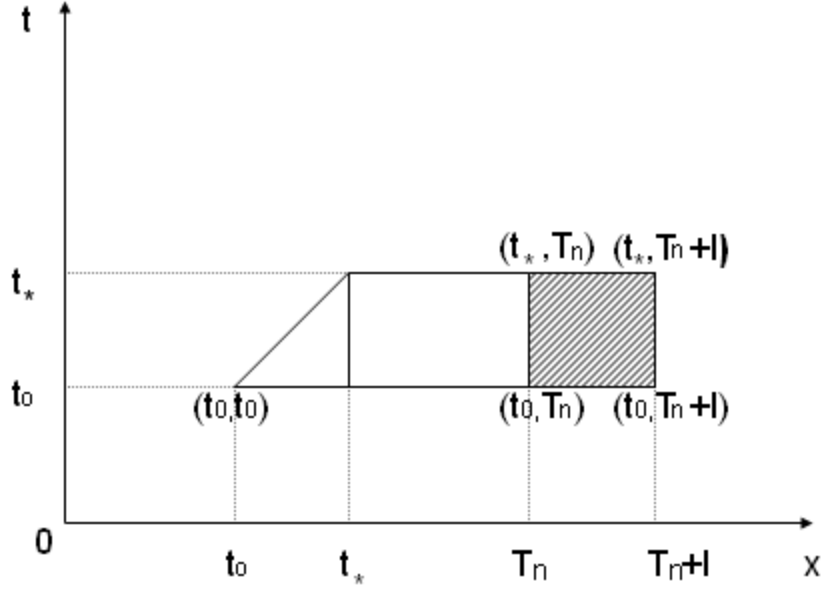, width=14cm}
  \caption{The domain of integration ${\cal M}$ for evaluating
  the drift of the Libor market numeraire. }
  \label{doml}
\end{figure}

In terms of the (compounded) forward interest rate Libor is given by
\begin{eqnarray}
L(t,T_n)=\frac{1}{\ell}\big(e^{\int^{T_n+\ell}_{T_n}dx
f(t,x)}-1\big)
\end{eqnarray}
To understand the discounting that yields a martingale evolution of
Libor rates $L(t_0,T_n)$ re-write Libor as follows
\begin{eqnarray}
L(t,T_n)&=&\frac{1}{\ell}\big(e^{\int^{T_n+\ell}_{T_n}dx
f(t,x)}-1\big)\nonumber\\
\label{liborfrbond}
          &=&\frac{1}{\ell}\big[\frac{B(t,T_n)-
          B(t,T_n+\ell)}{B(t,T_n+\ell)}\big]
\end{eqnarray}
The Libor is interpreted as being equal to $\big(B(t,T_n)-
B(t,T_n+\ell)\big)/l$, with the discounting factor for the Libor
market measure being equal to $B(t,T_n+\ell)$. Hence, the martingale
condition for the market measure, denoted by $E_L[..]$, is given by
\begin{eqnarray}
\frac{B(t_0,T_n)-
          B(t_0,T_n+\ell)}{B(t_0,T_n+\ell)}=E_L\Big[\frac{B(t_*,T_n)-
          B(t_*,T_n+\ell)}{B(t_*,T_n+\ell)}\Big]
\end{eqnarray}
In other words, the market measure is defined such that each Libor
is a martingale; that is, for $t_*>t_0$
\begin{eqnarray}
\label{mrktmt}
 L(t_0,T_n)=E_L[L(t_*,T_n)]
\end{eqnarray}
In terms of the underlying forward interest rates, the Libor's are
given by the following
\begin{eqnarray}
\label{fzfstr} F_0&\equiv& \int^{T_n+l}_{T_n}dx
f(t_0,x)~~;~~F_*\equiv \int^{T_n+l}_{T_n}dx f(t_*,x)\\
\label{mrktmt2}
 \Rightarrow L(t_0,T_n)&=&
\frac{1}{\ell}\big(e^{F_0}-1\big)~~;~~L(t_*,T_n)=
\frac{1}{\ell}\big(e^{F_*}-1\big)
\end{eqnarray}
and hence from eqs. \ref{mrktmt} and \ref{mrktmt2} the martingale
condition for Libor can be written as
\begin{eqnarray}
\label{mrtnmkt1}
 e^{F_0}=E_L[e^{F_*}]
\end{eqnarray}
Denote the drift for the market measure by $\alpha_L(t,x)$, and let
$T_n \leq x< T_n+\ell$; the evolution equation for the Libor forward
interest rates is given, similar to eq. \ref{frcfrwd}, by
\begin{eqnarray}
\label{frcmrkt}
 f(t,x)=f(t_0,x)+\int_{t_0}^t dt' \alpha_L(t',x)+\int_{t_0}^t
dt'\sigma(t',x)A(t',x)
\end{eqnarray}
Hence
\begin{eqnarray}
\label{mrtnmkt2} E_L\big[e^{F_*}\big]=e^{F_0+\int_{\cal M}
\alpha_L(t',x)} \frac{1}{Z} \int DA e^{\int_{\cal M}
\sigma(t',x)A(t',x)}e^{S[A]}
\end{eqnarray}
where the integration domain ${\cal M}$ is given in Fig. \ref{doml}.

Hence, from from eqs. \ref{zj}, \ref{mrtnmkt1} and \ref{mrtnmkt2}
\begin{eqnarray}
e^{-\int_{\cal M} \alpha_L(t,x)}&=&\int DA e^{\int_{\cal M}
\sigma(t,x)A(t,x)}e^{S[A]}\nonumber\\
&=&\exp\{\frac{1}{2}\int_{t_0}^{t_*}dt \int_{T_n}^{T_n+\ell} dx
dx'\sigma(t,x)D(x,x';t)\sigma(t,x')\}
\end{eqnarray}
Hence the Libor drift velocity is given by
\begin{eqnarray}
\alpha_L(t,x)=-\sigma(t,x)\int_{T_n}^x dx'
D(x,x';t)\sigma(t,x')~~;~~T_n \leq x < T_n+\ell
\end{eqnarray}
The Libor drift velocity $\alpha_L(t,x)$ is \textbf{negative}, as is
required for compensating growing payments due to the compounding of
interest.

There is a discontinuity in the value of $\alpha_L(t,x)$ at forward
time $x=T_n$; from its definition
\begin{eqnarray}
\alpha_L(t,T_n)=0
\end{eqnarray}
Approaching the value $\alpha_L(t,x)$ from $x>T_n$, the
discontinuity is given by
\begin{eqnarray}
\Delta \alpha_L(t,T_n)&\equiv&\lim_{x\rightarrow
T_n+}\alpha_L(t,T_n)-\alpha_L(t,T_n)\nonumber\\
      &=&-\sigma(t,x)\int_{T_n-\ell}^{T_n} dx'
D(x,x';t)\sigma(t,x')
\end{eqnarray}

Since the time-interval for Libor $\ell=90$ days is quite small, one
can approximate the drift by the following
\begin{eqnarray}
\label{driftapp} \alpha_L(t,x)\simeq -(x-T_n)\sigma^2(t,x) ~~;~~T_n
\leq x < T_n+\ell
\end{eqnarray}
since the normalization of the volatility function can always be
chosen so that $D(x,x;t)=1$ \cite{Baaquiecup}. The value of
discontinuity at $x=T_n$ is then approximately given by $-\ell
\sigma^2(t,T_n)$

\begin{figure}[h]
  \centering
  \epsfig{file=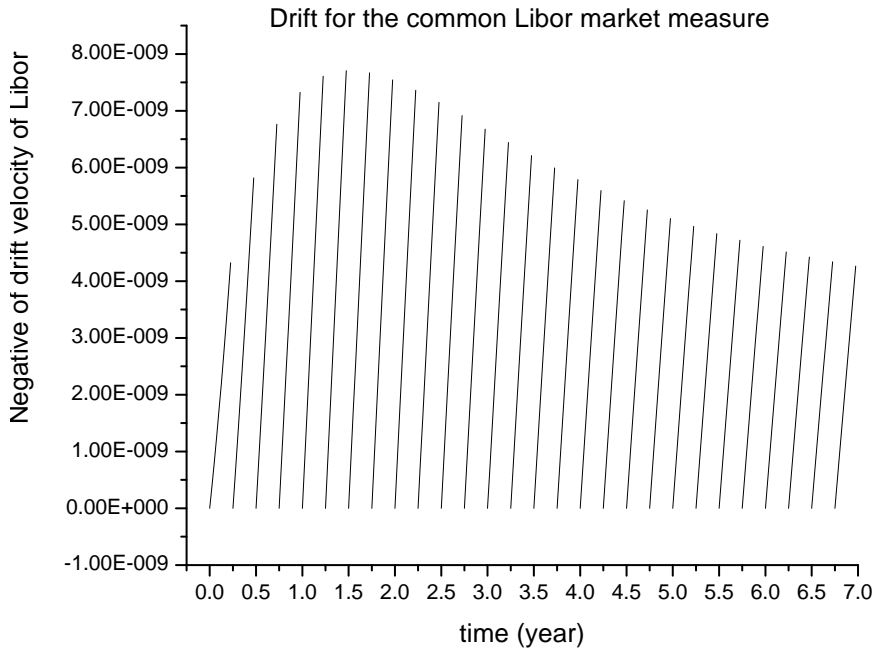, width=10cm, angle=-90}
  \caption{Negative of the drift velocity, namely $-\alpha_L(t,x)$, for the
  common Libor market measure, which is equal to the drift velocity $\alpha_F(t,x)$ for
  the forward Libor measure}
  \label{drift}
\end{figure}

Fig. \ref{drift} shows the behaviour of the drift velocity
$\alpha_F(t,x)$, with the value of $\sigma(t,x)$ taken from the
market \cite{Baaquiecup},\cite{data1}. One can see from the graph
that, in a given Libor interval, the drift velocity is approximately
linear in forward time and the maximum drift goes as
$\sigma^2(t,x)$, both of which is expected from eq. \ref{driftapp}.

\subsection{Libor Forward Measure}
It is often convenient to have a discounting factor that renders the
futures price of Libor Bonds into a martingale. Consider the Libor
forward bond given by
\begin{eqnarray}
\label{marfor0} F_L(t_0,T_{n+1})= e^{-\int_{T_n}^{T_{n+1}}dx
f(t_0,x)}=\frac{B(t_0,T_{n+1})}{B(t_0,T_{n})}
\end{eqnarray}
The forward numeraire is given by $B(t_0,T_{n})$; the drift velocity
is fixed so that the future price of a Libor bond is equal to its
forward value; hence
\begin{eqnarray}
\label{marfor1}
e^{-\int_{T_n}^{T_{n+1}}dx
f(t_0,x)}=E_F\big[e^{-\int_{T_n}^{T_{n+1}} dx f(t_*,x)}\big]
\end{eqnarray}

In effect, as expressed in the equation above, the forward measure
makes the forward Libor bond price a martingale. To determine the
corresponding drift velocity $\alpha_F(t,x)$, the right side of eq.
\ref{marfor1} is explicitly evaluated. Note from eq. \ref{frcfrwd}
\begin{eqnarray}
\label{marfor2} E_F\big[e^{-\int_{T_n}^{T_{n+1}}dx
f(t_*,x)}\big]&=&e^{-\int_{T_n}^{T_{n+1}}dx f(t_0,x)-\int_{\cal M}
\alpha_F(t',x)} \int DA e^{-\int_{\cal M}
\sigma(t',x)A(t',x)}e^{S[A]}\nonumber
\end{eqnarray}
where the integration domain ${\cal M}$ is given in Fig. \ref{doml}.

Hence, from eqs. \ref{zj}, \ref{marfor1} and \ref{marfor2}
\begin{eqnarray}
e^{\int_{\cal M} \alpha_F(t,x)}&=&\int DA e^{-\int_{\cal M}
\sigma(t,x)A(t,x)}e^{S[A]}\nonumber\\
  &=&\exp\{\frac{1}{2}\int_{t_0}^{t_*}dt \int_{T_n}^{T_{n+1}} dx
  dx'\sigma(t,x)D(x,x';t)\sigma(t,x')\}
\end{eqnarray}
Hence the drift velocity for the forward measure is given by
\begin{eqnarray}
\alpha_F(t,x)=\sigma(t,x)\int_{T_n}^x dx'
D(x,x';t)\sigma(t,x')~~;~~T_n \leq x < T_n+\ell
\end{eqnarray}

The Libor drift velocity $\alpha_L(t,x)$  is the negative of the
drift for the forward measure, that is
\begin{eqnarray}
\alpha_F(t,x)=-\alpha_L(t,x)\nonumber
\end{eqnarray}
Fig. \ref{drift} shows the behaviour of the drift velocity
$\alpha_F(t,x)$.

\subsection{Money Market Measure} In Heath, Jarrow,
and Morton \cite{HJM}, the martingale measure was defined by
discounting Treasury Bonds using the money market account, with the
money market numeraire $M(t,t_*)$ defined by
\begin{equation}
M(t,t_*)=e^{\int_{t}^{t_*}r(t')dt'} \, ,
\end{equation}
for the spot rate of interest $r(t)=f(t,t)$. The quantity
$B(t,T)/M(t,t)$ is defined to be a martingale
\begin{eqnarray}
\frac{B(t,T)}{M(t,t)}&=&E_M \big[\frac{B(t_*,T)}{M(t,t_*)}\big]  \nonumber\\
\label{mmmar}
 \Rightarrow B(t,T)&=&E_M\big[e^{-\int_{t}^{t_*}r(t')dt'}B(t_*,T)\big]
\end{eqnarray}
where $E_M[..]$ denotes expectation values taken with respect to the
money market measure. The martingale condition can be solved for
it's corresponding drift velocity, which is given by
\begin{eqnarray}
\alpha_M(t,x)=\sigma(t,x)\int_t^x dx'D(x,x';t)\sigma(t,x')
\end{eqnarray}

\section{Pricing a
 Mid-Curve Cap} \label{pricing}
An interest rate cap is composed out of a linear sum of individual
caplets. The pricing formula for an interest rate caplet is obtained
for a general volatility function $\sigma(t,x)$ and propagator
$D(x,x^{'};t)$ that drive the underlying Libor forward rates.

\begin{figure}[h]
  \centering
  \epsfig{file=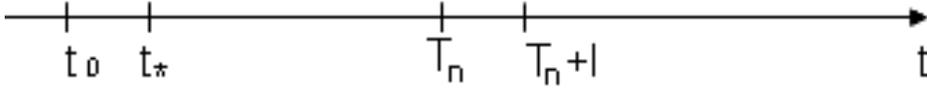, width=14cm}
  \caption{Time intervals in the pricing of a caplet. }
  \label{times}
\end{figure}

A mid-curve caplet can be exercised at any fixed time $t_*$ that is
less then the time $T_n$ at which the caplet matures. Denote by
$Caplet(t_0,t_*,T_n)$ the price -- at time $t_0$ --  of an interest
rate European option contract that must be exercised at time
$t_*>t_0$ for an interest rate caplet that puts an upper limit to
the interest from time $T_n$ to $T_n+\ell$. Let the principal amount
be equal to $\ell V$, and the caplet rate be $K$. The caplet is
exercised at time $t_*$, with the payment made in arrears at time
$T_n+\ell$. Note that although the payment is made at time
$T_n+\ell$, the \textbf{amount} that will be paid is fixed at time
$t_*$. The various time intervals that define the interest rate
caplet are shown in Fig.\ref{times}.

The payoff function of an interest rate caplet is the value of the
caplet when it matures, at $t_0=t_*$, and is given by
\begin{eqnarray}
\label{payoffcaplet2}
 Caplet(t_*,t_*,T_n)&=&\ell V  B(t_*,T_n+\ell)\big[L(t_*,T_n)-K\big]_+ \\
          &=&\ell V \big[\frac{B(t_*,T_n)-
          B(t_*,T_n+\ell)}{\ell}-KB(t_*,T_n+\ell)\big]_+ \nonumber\\
\label{payoffcaplet}
          &=&\tilde{V} B(t_*,T_n+\ell)\big(Xe^{F_*}-1 \big)_+
 \end{eqnarray}
 where recall from eq. \ref{marfor2}
 \begin{eqnarray}
  F_*\equiv\int_T^{T_n+\ell} dx
          f(t_*,x)~~\mathrm{and}~~X=\frac{1}{1+\ell{K}}~~;~~\tilde{V}=(1+\ell{K})V
          \nonumber
\end{eqnarray}
The payoff for an interest rate floorlet is similarly given by
\begin{eqnarray}
Floorlet(t_*,t_*,T_n)&=&\ell V
B(t_*,T_n+\ell)\big[K-L(t_*,T_n)\big]_+
          \nonumber\\
\label{payofffloor}
          &=&\tilde{V} B(t_*,T_n+\ell)\big(1-Xe^{F_*} \big)_+
 \end{eqnarray}
As will be shown in Section \ref{pcsec}, the price of the caplet
automatically determines the price of a floorlet due to put-call
parity, and hence the price of the floorlet does not need an
independent derivation.

An interest rate cap of a duration over a longer period is made from
the sum over the caplets spanning the requisite time interval.
Consider a mid-curve cap, to be exercised at time $t_*$, with strike
price $K_j$ from time $j\ell$ to time $j(+1)\ell$, and with the
interest cap starting from time $T_m=m\ell$ and ending at time
$(n+1)\ell$; its price is given by
\begin{eqnarray}
\label{capprice}
Cap(t_0,t_*)=\sum_{j=m}^n  Caplet(t_0,t_*,T_j;K_j)
\end{eqnarray}
and a similar expression for an interest rate floor in terms of the
floorlets for a single Libor interval.
\subsection{Forward Measure Calculation for Caplet}
The numeraire for the forward measure is given by the Libor Bond
$B(t,T_n)$. Hence the caplet is a martingale when discounted by
$B(t,T_n)$; the price of the caplet at time $t_0<t_*$ is
consequently given by
\begin{eqnarray}
 \frac{Caplet(t_0,t_*,T_n)}{B(t_0,T_n)}&=&
E_F\big[\frac{Caplet(t_*,t_*,T_n)}{B(t_*,T_n)}\big]\nonumber\\
 &=& \tilde{V}E_F\big(X-e^{-F_*}\big)_+\nonumber
\end{eqnarray}
Hence, in agreement with eq. \ref{payoffcaplet}, the price of a
caplet is given by
\begin{eqnarray}
\label{capletfm1}
 Caplet(t_0,t_*,T_n)=\tilde{V}B(t_0,T_n)
E_F\big(X-e^{-F_*}\big)_+
\end{eqnarray}
The payoff function for the caplet given in eq. \ref{capletfm1}
above for the interest caplet has been obtained in \cite{Baaquiecup}
and \cite{JTtext} using a different approach.

The price of the caplet is given by \cite{Baaquiecup}
\begin{eqnarray}
\label{capletfm} Caplet(t_0,t_*,T_n)= \tilde{V}B(t_0,T_n)
\int_{-\infty}^{+\infty}d{G}\Psi_F(G)(X-e^{-G})_{+}
\end{eqnarray}
From the derivation given in \cite{Baaquiecup}, the pricing kernel
$\Psi_F(G)=\Psi_F(G,t_0,t_*,T_n)$ is given by
\begin{eqnarray}
\label{kerfm}
 \Psi_F(G)&=&\sqrt{\frac{1}{2\pi{q}^{2}}}
\exp\left\{-\frac{1}{2q^{2}}
\left(G-\int_{T_n}^{T_n+\ell}dx{f}(t_0,x)-\frac{q^{2}}{2}\right)^{2}\right\}\\
\label{qsqr} q^{2}&=&q^2(t_0,t_*,T_n)\nonumber\\
&=&\int_{t_0}^{t_*}dt\int_{T_n}^{T_n+\ell}dx{d}x^{'}
\sigma(t,x)D(x,x^{'};t)\sigma(t,x^{'})\nonumber
\end{eqnarray}

The price of the caplet is given by the following Black-Scholes type
formula
\begin{eqnarray}
Caplet(t_0,t_*,T_n)=\tilde{V}B(t_0,T)\left[X{N}(-d_{-})-{\cal
F}N(-d_{+})\right]
\end{eqnarray}
where $N(d_{\pm})$ is  the cumulative distribution for the normal
random variable with the following definitions
\begin{eqnarray}
\label{qcaplet}
{\cal F}&=&e^{-\int_{T_n}^{T_n+\ell}dx{f}(t_0,x)}=e^{-F_0}\nonumber\\
d_{\pm}&=&\frac{1}{q}\left[\ln\left(\frac{{\cal
F}}{X}\right)\pm\frac{q^{2}}{2}\right] \
\end{eqnarray}

\subsection{Libor Market Measure Calculation for Caplet}
The Libor market measure has as its numeraire the Libor bond
$B(t_*,T_n+\ell)$; the caplet is a martingale when discounted by
this numeraire, and hence the price of the caplet at time $t_0<t_*$
is given by
\begin{eqnarray}
\frac{Caplet(t_0,t_*,T_n)}{B(t_0,T_n+\ell)}&=&
E_L\big[\frac{Caplet(t_*,t_*,T_n)}{B(t_*,T_n+\ell)}\big]\nonumber\\
&=& \tilde{V}E_L\big(Xe^{F_*}-1\big)_+\nonumber \\
\label{capletlmm} \Rightarrow
Caplet(t_0,t_*,T_n)&=&\tilde{V}B(t_0,T_n+\ell)E_L\big(Xe^{F_*}-1\big)_+
\end{eqnarray}
where, similar to the derivation given in \cite{Baaquiecup}, the
price of the caplet is given by
\begin{eqnarray}
\label{pricecapl}
Caplet(t_0,t_*,T_n)&=&\tilde{V}B(t_0,T_n+\ell)\int_{-\infty}^{+\infty}d{G}\Psi_L(G)(Xe^{G}-1)_{+}
\end{eqnarray}
For $\Psi_L(G)=\Psi_L(G,t_0,t_*,T_n)$ the pricing kernel is given by
\begin{eqnarray}
\label{psil}
 \Psi_L(G)= \sqrt{\frac{1}{2\pi{q}^{2}}}
\exp\left\{-\frac{1}{2q^{2}}
\left(G-\int_{T_n}^{T_n+\ell}dx{f}(t_0,x)+\frac{q^{2}}{2}\right)^{2}\right\}
\end{eqnarray}

The price of the caplet obtained from the forward measure is equal
to the one obtained using the Libor market measure since, from eqs.
\ref{capletfm} and \ref{kerfm}, one can prove the following
remarkable result
\begin{eqnarray}
B(t,T_n) \Psi_F(G)(X-e^{-G})_{+}
=B(t,T_n+\ell)\Psi_L(G)(Xe^{G}-1)_{+}
\end{eqnarray}
The identity above shows how the three factors required in the
pricing of an interest rate caplet, namely the discount factors, the
drift velocities and the  payoff functions, all `conspire' to yield
numeraire invariance for the price of the interest rate option.

The payoff function is correctly given by the price of the caplet,
since in the limit of $t_0\rightarrow t_*$,  eq. \ref{qsqr} yields
\begin{eqnarray}
\lim_{t_0\rightarrow
t_*}q^{2}&=&(t_*-t_0)\int_{T_n}^{T_n+\ell}dx{d}x^{'}
\sigma(t,x)D(x,x^{'};t)\sigma(t,x^{'})\nonumber \\
     &=& \epsilon C
\end{eqnarray}
where $C$ is a constant, and $\epsilon=t_*-t_0$. Hence, from eqs.
\ref{pricecapl} and \ref{psil}
\begin{eqnarray}
\lim_{t_0\rightarrow
t_*}Caplet(t_0,t_*,T_n)&=&\tilde{V}B(t_*,T_n+\ell)\int_{-\infty}^{+\infty}d{G}\delta(G-F_*)(Xe^{G}-1)_{+}\nonumber\\
                            &=&\tilde{V}B(t_*,T_n+\ell)(Xe^{F_*}-1)_{+}\nonumber
\end{eqnarray}
verifying the payoff function as given in eq. \ref{payoffcaplet}.

\subsection{Money Market Calculation for Caplet}
\begin{figure}[h]
  \centering
  \epsfig{file=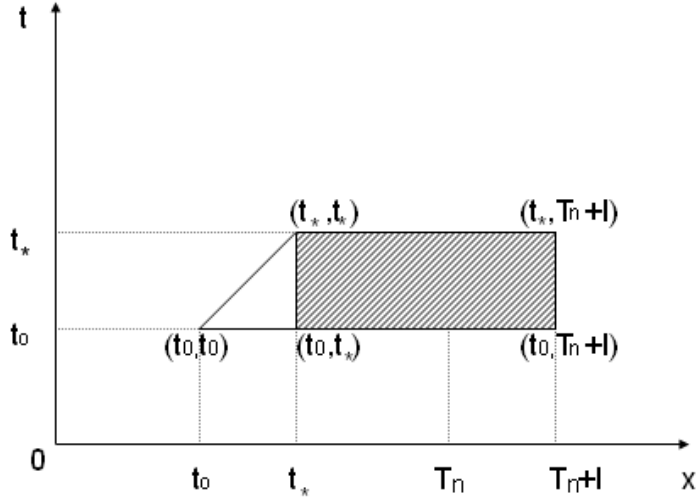, width=12cm, angle=0}
  \caption{Domain of integration ${\cal R}$ for evaluating the price of a
  caplet using the money market numeraire. }
  \label{domr}
\end{figure}

The money market numeraire is given by the spot interest rate
$M(t_0,t_*)=exp\{\int_{t_0}^{t_*}dt r(t)\}$. Expressed in terms of
the money martingale numeraire, the price of the caplet is given by
\begin{eqnarray}
\frac{Caplet(t_0,t_*,T_n)}{M(t_0,t_0)}&=&E_M\big[\frac{Caplet(t_*,t_*,T_n)}{M(t_0,t_*)}\big]\nonumber\\
\Rightarrow Caplet(t_0,t_*,T_n)&=&E_M\big[e^{-\int_{t_0}^{t_*}dt
r(t)}Caplet(t_*,t_*,T_n)\big]\nonumber
\end{eqnarray}
To simplify the calculation, consider the change of numeraire from
$M(t_0,t_*)=exp\{\int_{t_0}^{t_*}dt' r(t')\}$ to discounting by the
Treasury Bond $B(t_0,t_*)$; it then follows \cite{Baaquiecup} that
\begin{eqnarray}
e^{-\int_{t_0}^{t_*}dt r(t)} e^S=B(t_0,t_*)e^{S_*}\nonumber
\end{eqnarray}
where the drift for the action $S_*$ is given by
\begin{eqnarray}
\label{alpstr}
 \alpha_*(t,x)=\sigma(t,x)\int_{t_*}^x
dx'D(x,x';t)\sigma(t,x')
\end{eqnarray}
In terms of the money market measure, the price of the caplet is
given by
\begin{eqnarray}
\label{capletmm}
 Caplet(t_0,t_*,T_n)&=&E_M\big[e^{-\int_{t_0}^{t_*}dt
r(t)}Caplet(t_*,t_*,T_n)\big]\\
         &=&B(t_0,t_*)E_M^*\big[Caplet(t_*,t_*,T_n)\big]\nonumber\\
          &=&\tilde{V}B(t_0,t_*) E_M^*\big[B(t_*,T_n+\ell)\big(Xe^{F_*}-
          1\big)_+\big]\nonumber
\end{eqnarray}
From the expression for the forward rates given in eq. \ref{frcfrwd}
the price of the caplet can be written out as follows
\begin{eqnarray}
\label{mnymrft}
 Caplet(t_0,t_*,T_n)&=&\tilde{V} B(t_0,t_*)
E_M^*\big[B(t_*,T_n+\ell)\big(Xe^{F_*}-
          1 \big)_+\big]\nonumber\\
          &=& \tilde{V}B(t_0,T_n+\ell)e^{-\int_{\cal R}\alpha_*}~
          \frac{1}{Z}\int DA e^{-\int_{\cal R}\sigma A}e^{S_*}  \big(Xe^{F_*}-1 \big)_+
\end{eqnarray}
where the integration domain ${\cal R}$ is given in Fig. \ref{domr}.

The payoff can be re-expressed using the Dirac delta-function as
follows
\begin{eqnarray}
\label{drdelfn}
\big(Xe^{F_*}-1 \big)_+ &=&\int dG \delta(G-F_*)\big(Xe^{G}-1\big)_+ \nonumber\\
          &=&\int dG \int \frac{d\xi}{2\pi} e^{i\xi(G-F_*)}\big(Xe^{G}-1 \big)_+
\end{eqnarray}
From eq. \ref{frcfrwd}, and domain of integration ${\cal M}$ given
in Fig. \ref{doml}, one obtains
\begin{eqnarray}
F_*&\equiv&\int_{T_n}^{T_n+\ell}dx f(t_*,x)\nonumber\\
    &=&\int_{T_n}^{T_n+\ell}dx
f(t_0,x)+\int_{{\cal M}} \alpha_*+\int_{{\cal M}}\sigma A \nonumber
\end{eqnarray}

Hence, from eqs.\ref{mnymrft} and \ref{drdelfn} the price of the
caplet, for $F_0=\int_{T_n}^{T_n+\ell}dx f(t_0,x)$, is given by
\begin{eqnarray}
\label{capletmm} Caplet(t_0,t_*,T_n)&=& \tilde{V}B(t_0,t_*)
E_M^*\big[B(t_*,T_n+\ell)\big(Xe^{F_*}-
          1 \big)_+\big]\nonumber\\
          &=&\tilde{V} B(t_0,T_n+\ell)e^{-\int_{\cal R}\alpha_*}~\int dG \int \frac{d\xi}{2\pi}
          e^{i\xi(G-F_0-\int_{{\cal M}} \alpha_*)}\big(Xe^{G}-1 \big)_+ \times \nonumber\\
          &&\frac{1}{Z}\int DA e^{-\int_{\cal R}\sigma A}e^{-i\xi\int_{{\cal M}}\sigma A} e^{S_*}
\end{eqnarray}
To perform  path integral note that
\begin{eqnarray}
\int_{\cal R}\sigma A+i\xi\int_{{\cal M}}\sigma A=\int_{t_0}^{t_*}dt
\int_{t_*}^{T_n+\ell} dx \sigma(t,x)A(t,x)+i\xi\int_{t_0}^{t_*}dt
\int_{T_n}^{T_n+\ell} dx \sigma(t,x)A(t,x)\nonumber
\end{eqnarray}
and the Gaussian path integral using eq. \ref{zj} yields
\begin{eqnarray}
\frac{1}{Z}\int DA e^{-\int_{\cal R}\sigma A-i\xi\int_{{\cal
R}_L}\sigma A} e^{S_*}=e^\Gamma \nonumber
\end{eqnarray}
where
\begin{eqnarray}
 &&\Gamma
=\frac{1}{2}\int_{t_0}^{t_*}dt\int_{t_*}^{T_n+\ell} dx
dx'\sigma(t,x)D(x,x';t)\sigma(t,x')\nonumber\\
&&~~~~~-\frac{\xi^2}{2}\int_{t_0}^{t_*}dt\int_{T_n}^{T_n+\ell} dx
dx'\sigma(t,x)D(x,x';t)\sigma(t,x')\nonumber\\
&&~~~~~~~~~~+i\xi\int_{t_0}^{t_*}dt\int_{t_*}^{T_n+\ell} dx
\int_{T_n}^{T_n+\ell}dx'\sigma(t,x)D(x,x';t)\sigma(t,x')\nonumber
\end{eqnarray}
The expression for $\Gamma$ above, using the definition of $ q^2,
\alpha_*$ given in eqs. \ref{qsqr} and \ref{alpstr} respectively,
can be shown to yield the following
\begin{eqnarray}
\label{gamma} \Gamma=\int_{\cal
R}\alpha_*-\frac{\xi^2}{2}q^2+i\xi\big(\int_{{\cal M}}
\alpha_*+\frac{1}{2}q^2\big)
\end{eqnarray}
Simplifying eq. \ref{capletmm} using eq. \ref{gamma} yields the
price of the caplet as given by
\begin{eqnarray}
 Caplet(t_0,t_*,T_n) =\tilde{V}B(t_0,T_n+\ell)
\int_{-\infty}^{+\infty}d{G}\Psi_L(G)(Xe^{G}-1)_{+}
\end{eqnarray}
Hence we see that the money market numeraire yields the same price
for the caplet as the ones obtained from the forward and Libor
market measure, as expected, but with a derivation that is very
different from the previous ones.

\section{Put-Call Parity for Caplets and Floorlets} \label{pcsec}
Put-call parity for caplets and floorlets is a model independent
result, and is derived by demanding that the prices be equal of two
portfolios -- having identical payoffs at maturity -- formed out of
a caplet and the money market account on the one hand, and a
floorlet and futures contract on the other hand \cite{JTtext}.
Failure of the prices to obey the put-call parity relation would
then lead to arbitrage opportunities. More precisely, put-call
parity yields the following relation between the price of a caplet
and a floorlet
\begin{eqnarray}
\label{pcmip}
Caplet(t_0,t_*,T_n)+\tilde{V}B(t_0,T_n+\ell)=Floorlet(t_0,t_*,T_n)+
\tilde{V}B(t_0,T_n+\ell)Xe^{F_0}
\end{eqnarray}
where the other two instruments are the money market account  and a
futures contract.

Re-arranging eq. \ref{pcmip} and simplifying yields
\begin{eqnarray}
\label{capfllt}
Caplet(t_0,t_*,T_n)-Floorlet(t_0,t_*,T_n)&=& \ell V
B(t_0,T_n+\ell)[L(t_0,T_n)-K] \\
               &=&\mathrm{value~of~swaplet}\nonumber
\end{eqnarray}
The right hand side of above equation is the price, at time $t_0$,
of a forward or deferred swaplet, which is an interest rate swaplet
that matures at time $T_n$; swaps are discussed in Section
\ref{swaps}.

In this Section a derivation is given for put-call parity for
(Libor) options ; the derivation is given for the three different
numeraires, and illustrates how the properties are essential for the
numeraires to price the caplet and floor so that they satisfy
put-call parity.

The payoff for the caplet and a floorlet is generically given by
\begin{eqnarray}
(a-b)_+=(a-b)\Theta(a-b)\nonumber
\end{eqnarray}
where the Heaviside step function $\Theta(x)$ is defined by
\begin {eqnarray}
\label{thetafn} \Theta(x) = \left\{ \begin{array}{ll}
1 & x > 0 \\
\frac{1}{2} & x = 0 \\
0 & x < 0
\end{array}
\right. \nonumber
\end {eqnarray}
The derivation of put-call parity hinges on the following identity
\begin{equation}
\label{theta} \Theta(x)+\Theta(-x)=1
\end{equation}
since it yields, for the difference of the payoff functions of the
put and call options, the following
\begin{eqnarray}
\label{pcparity}
(a-b)_+-(b-a)_+&=&(a-b)\Theta(a-b)-(b-a)\Theta(b-a)\nonumber\\
               &=&a-b
\end{eqnarray}

\subsection{Put-Call Parity for Forward Measure}
The price of a caplet and floorlet at time $t_0$ is given by
discounting the payoff functions with the discounting factor of
$B(t_0,T_n)$. From eq. \ref{capletfm}
\begin{eqnarray}
Caplet(t_0,t_*,T_n)&=&B(t_0,T_n)
E_F\big[\frac{Caplet(t_*,t_*,T_n)}{B(t_*,T_n)}\big]\nonumber\\
     &=& \tilde{V}B(t_0,T_n)
E_F\big(X-e^{-F_*}\big)_+\nonumber
\end{eqnarray}
and the floorlet is given by
\begin{eqnarray}
\label{flrfm}
Floorlet(t_0,t_*,T_n)=\tilde{V}B(t_0,T_n)
E_F\big(e^{-F_*}-X\big)_+
\end{eqnarray}
Consider the expression
\begin{eqnarray}
&&Caplet(t_0,t_*,T_n)-Floorlet(t_0,t_*,T_n)\\
&=&\tilde{V}B(t_0,T_n)\Big[E_F\big(X-e^{-F_*}\big)_+
- E_F\big(e^{-F_*}-X\big)_+ \Big]\nonumber\\
\label{pcfm}
  &=&\tilde{V}B(t_0,T_n)E_F\Big(X-e^{-F_*}\Big)
\end{eqnarray}
where eq. \ref{pcparity} has been used to obtain eq. \ref{pcfm}.

For the forward measure, from eq. \ref{marfor1}
\begin{eqnarray}
E_F\big[e^{-F_*}\big]=e^{-F_0}
\end{eqnarray}
Hence, since for constant $X$ we have $E_F(X)=XE_F(1)=X$, from above
equation and eq. \ref{pcfm}, the price of a caplet and floorlet
obeys the put-call relation
\begin{eqnarray}
\label{pcpcpflt}
Caplet(t_0,t_*,T_n)-Floorlet(t_0,t_*,T_n)&=&\tilde{V}B(t_0,T_n)E_F\Big(X-e^{-F_*}\Big)\nonumber\\
                                 &=&\tilde{V}B(t_0,T_n)(X-e^{-F_0})\nonumber\\
                                 &=&\ell VB(t_0,T_n+\ell)(L(t_0,T_n)-K)
\end{eqnarray}
and yields eq. \ref{pcmip} as expected.
\subsection{Put-Call for Libor Market Measure}
The price of a caplet for the Libor market measure is given from eq.
\ref{capletlmm} by
\begin{eqnarray}
Caplet(t_0,t_*,T_n)=\tilde{V}B(t_0,T_n+\ell)E_L\big(Xe^{F_*}-1\big)_+
\end{eqnarray}
and the floorlet is given by
\begin{eqnarray}
Floorlet(t_0,t_*,T_n)=\tilde{V}B(t_0,T_n+\ell)E_L\big(1-Xe^{F_*}\big)_+
\end{eqnarray}
Hence, similar to the derivation given in eq.\ref{pcfm}, we have
\begin{eqnarray}
\label{pclmm1}
Caplet(t_0,t_*,T_n)-Floorlet(t_0,t_*,T_n)=\tilde{V}B(t_0,T_n+\ell)E_L\big(Xe^{F_*}-1\big)
\end{eqnarray}
For the Libor market measure, from eq.\ref{mrtnmkt1}
\begin{eqnarray}
E_L[e^{F_*}]=e^{F_0} \nonumber
\end{eqnarray}
and hence equation above, together with eq. \ref{pclmm1}, yields the
expected eq. \ref{pcmip} put-call parity relation
\begin{eqnarray}
Caplet(t_0,t_*,T_n)-Floorlet(t_0,t_*,T_n)&=&\tilde{V}B(t_0,T_n+\ell)(Xe^{F_0}-1)
\nonumber\\
  &=&\ell VB(t_0,T_n+\ell)(L(t_0,T_n)-K)\nonumber
\end{eqnarray}
\subsection{Put-Call for Money Market Measure}
The money market measure has some interesting intermediate steps in
the derivation of put-call parity. Recall the caplet for the money
market measure is given from eq. \ref{capletmm} as
\begin{eqnarray}
Caplet(t_0,t_*,T_n)&=&E_M\big[e^{-\int_{t_0}^{t_*}dt
r(t)}Caplet(t_*,t_*,T_n)\big]\nonumber
\end{eqnarray}
Using the definition of the payoff function for a caplet given in
eq. \ref{payoffcaplet} yields
\begin{eqnarray}
Caplet(t_0,t_*,T_n)=\tilde{V}E_M\Big (e^{-\int_{t_0}^{t_*}dt
r(t)}\big[X B(t_*,T_n)-
          B(t_*,T_n+\ell)\big]_+\Big) \nonumber
\end{eqnarray}
The price of the floor is given by
\begin{eqnarray}
Floorlet(t_0,t_*,T_n)=\tilde{V}E_M\Big(e^{-\int_{t_0}^{t_*}dt
r(t)}\big[B(t_*,T_n+\ell)-X B(t_*,T_n)
          \big]_+ \Big)\nonumber
\end{eqnarray}
Consider the difference of put and call on a caplet; similar to the
previous cases, using eq. \ref{theta} yields the following
\begin{eqnarray}
Caplet(t_0,t_*,T_n)-Floorlet(t_0,t_*,T_n)=\tilde{V}E_M\Big
(e^{-\int_{t_0}^{t_*}dt r(t)}\big[X B(t_*,T_n)-
          B(t_*,T_n+\ell)\big]\Big)
\end{eqnarray}
The martingale condition given in eq. \ref{mmmar} yields the
expected result given in eq. \ref{pcmip} that
\begin{eqnarray}
Caplet(t_0,t_*,T_n)-Floorlet(t_0,t_*,T_n)&=&\tilde{V}\big[X
B(t_0,T_n)-
          B(t_0,T_n+\ell)\big]\nonumber\\
                 &=&\tilde{V}B(t_0,T_n+\ell)(Xe^{F_0}-1)\nonumber\\
                 &=&\ell VB(t_0,T_n+\ell)(L(t_0,T_n)-K)\nonumber
\end{eqnarray}

To obtain  put-call parity for the money market account, unlike the
other  two cases, \textbf{two} instruments, namely
$e^{-\int_{t_0}^{t_*}dt r(t)}B(t_*,T_n)$ and $e^{-\int_{t_0}^{t_*}dt
r(t)}B(t_*,T_n+\ell)$, have to be martingales, which in fact turned
out to be the case for the money market numeraire.

\section{Swaps, Caps and Floors}\label{swaps}
An interest swap is contracted between two parties. Payments are
made at fixed intervals, usually $90$  or $180$ days, denoted by
$T_n$, with the contract having notional principal $\ell V$, and a
pre-fixed total duration, with the last payment made at time
$T_N+\ell$. A swap of the first kind, namely swap$_I$, is where one
party pays at a fixed interest rate $R_S$ on the notional principal,
and the other party pays a floating interest rate based on the
prevailing Libor rate. A swap of the second kind, namely
swap$_{II}$, is where the party pays at the floating Libor rate and
receives payments at fixed interest rate $R_S$.

To quantify the value of the swap, let the contract start at time
$T_0$, with payments made at fixed interval $T_n=T_0+n\ell$, with
times $n=0,1,2, ...,N$.

Consider a swap in which the payments at the fixed rate is given by
$R_S$; the values of the swaps are then given by \cite{JTtext}
\begin{eqnarray}
\label{swap12}
\mathrm{swap}_I(T_0, R_S)=V\big [1-B(T_0,T_N+\ell)-\ell R_S\sum_{n=0}^N B(T_0,T_n+\ell)\big ]\nonumber\\
\mathrm{swap}_{II}(T_0, R_S)=V\big [\ell R_S\sum_{n=0}^N
B(T_0,T_n+\ell)+B(t,T_N+\ell)-1\big]
\end{eqnarray}

The par value of the swap when it is initiated, that is at time
$T_0$, is zero; hence the par fixed rate $R_P$, from eq.
\ref{swap12}, is given by \cite{JTtext}
\begin{eqnarray}
\mathrm{swap}_I(T_0, R_P)&=&0=\mathrm{swap}_{II}(T_0, R_P)\nonumber\\
\Rightarrow \ell R_P&=& \frac{1-B(T_0,T_N+\ell)}{\sum_{n=0}^N
B(T_0,T_n+\ell)}\nonumber
\end{eqnarray}

Recall from eqs. \ref{capfllt} and \ref{capprice} that a cap or a
floor is constructed from a linear sum of caplets and floorlets. The
put-call parity for interest rate caplets and floorlets given in eq.
\ref{capfllt} in turn yields
\begin{eqnarray}
\label{capflrprty}
Cap(t_0,t_*)-Floor(t_0,t_*)
&=&\sum_{j=m}^n \big [Caplet(t_0,t_*,T_j;K)-Floorlet(t_0,t_*,T_j;K)\big ]\nonumber\\
&=&\ell V\sum_{n=0}^N B(t_0,T_n+\ell)\big[L(t_0,T_n)-K\big]
\end{eqnarray}

The price of a swap at time $t_0<T_0$ is similar to the forward
price of a Treasury Bond, and is called a forward swap or a deferred
swap.\footnote{A swap that is entered into after the time of the
initial payments, that is, at time $t_0>T_0$ can also be priced and
is given in \cite{JTtext}; however, for the case of a swaption, this
case is not relevant.} Put-call parity for caps and floors gives the
value of a forward swap, and hence
\begin{eqnarray}
\label{swap1}
\mathrm{swap}_I(t_0, R_S)=\ell V\sum_{n=0}^N
B(t_0,T_n+\ell)\big[L(t,T_n)-R_S\big]\\
\label{swap2} \mathrm{swap}_{II}(t_0, R_S)=\ell V\sum_{n=0}^N
B(t_0,T_n+\ell)\big[R_S-L(t_0,T_n)\big]
\end{eqnarray}

The value of the swaps, from eqs. \ref{swap1} and \ref{swap2}, can
be seen to have the following intuitive interpretation: At time
$T_n$ the value of swap$_I$ is the difference between the floating
payment received at the rate of $L(t,T_n)$, and the fixed payments
paid out at the rate of $R_S$. All payments are made at time
$T_n+\ell$, and hence for obtaining its value at time $t_0$ need to
be discounted by the bond $B(t_0,T_n+\ell)$.

The definition of $L(t_0,T_n)$ given in eq. \ref{liborfrbond} yields
the following
\begin{eqnarray}
\ell V\sum_{n=0}^N B(t_0,T_n+\ell)L(t_0,T_n)&=& V \sum_{n=0}^N \big [B(t_0,T_n)-B(t_0,T_n+\ell)\big]\nonumber\\
                        &=&V \big [B(t_0,T_0)-B(t_0,T_N+\ell)\big]\nonumber\\
\label{swap1fl}
                        &\Rightarrow&V \big
                        [1-B(T_0,T_N+\ell)\big]~~\mathrm{for}~t_0=T_0
\end{eqnarray}

Hence, from eq. \ref{swap1}
\begin{eqnarray}
\label{swap11} \mathrm{swap}_{I}(t_0,R_S)=\ell
V\big[B(t_0,T_0)-B(t_0,T_N+\ell)-\ell R_S \sum_{n=0}^N
B(t_0,T_n+\ell)\big]
\end{eqnarray}
with a similar expression for $\mathrm{swap}_{II}$. Note that the
forward swap prices, for $t_0 \to T_0$,  converge to the expressions
for swaps given in eqs. \ref{swap1} and \ref{swap1}.

At time $t_0$ the par value for the fixed rate of the swap, namely
$R_P(t_0)$, is given by both the forward swaps being equal to zero.
Hence
\begin{eqnarray}
\mathrm{swap}_I(t_0, R_P(t_0))&=&0=\mathrm{swap}_{II}(t_0, R_P(t_0))\nonumber\\
\Rightarrow \ell R_P(t_0)&=&
\frac{B(t_0,T_0)-B(t_0,T_N+\ell)}{\sum_{n=0}^N
B(t_0,T_n+\ell)}\nonumber\\
&=& \frac{1-F(t_0,T_0,T_N+\ell)}{\sum_{n=0}^N
F(t_0,T_0,T_n+\ell)}~~~~\mathrm{where}~~~F(t_0,T_0,T_n+\ell)=e^{-\int_{T_0}^{T_n+\ell}dxf(t_0,x)}\nonumber\\
  \Rightarrow \lim_{t_0 \to T_0}R_P(t_0)&=&R_P
\end{eqnarray}
We have obtained the anticipated result that the par value for the
forward swap is fixed by the forward bond prices
$F(t_0,T_0,T_n+\ell)$, and converges to the par value of the swap
when it matures at time $t_0=T_0$.

In summary, put-call parity for cap and floor, from eqs.
\ref{capflrprty} and \ref{swap1} yields, for $K=R_S$
\begin{eqnarray}
Cap(t_*,t_*;R_S)-Floor(t_0,t_*;R_S)=\mathrm{swap}_I(t_0, R_S)
\end{eqnarray}
as expected \cite{JTtext}.
\section{Conclusions}
A common Libor market measure was derived, and it was shown that a
single numeraire renders all Libor into martingales. Two other
numeraires were studied for the forward interest rates, each having
its own drift velocity.

All the numeraires have their own specific advantages, and it was
demonstrated by actual computation that all three  yield the same
price for an interest rate caplet, and also satisfy put-call parity
as is necessary for the prices interest caps and floors to be free
from arbitrage opportunities.

The expression for the payoff function for the caplet given in eq.
\ref{payoffcaplet2}, namely
\begin{eqnarray}
Caplet(t_0,t_*,T_n)=\ell V
B(t_*,T_n+\ell)\big[L(t_*,T_n)-K\big]_+\nonumber
\end{eqnarray}
is seen to be the correct one as it reproduces the payoff functions
that are widely used in the literature, yields a pricing formula for
the interest rate caplet that is numeraire invariant, and satisfies
the requirement of put-call parity as well.

An analysis of swaps shows that put-call parity for caps and floors
correctly reproduces the swap future as expected.

\section{Acknowledgements}

I am greatly indebted to Sanjiv Das for sharing his insights of
Libor with me, and which were instrumental in clarifying this
subject to me. I would like to thank Mitch Warachka for a careful
reading of the manuscript, and to Cui Liang for many discussions.

\end{document}